\newcommand{\ket}[1]{\ensuremath{\left| {#1} \right>}}
\newcommand{\bra}[1]{\ensuremath{\left< {#1} \right|}}

\newcommand{\Be}{\ensuremath{{^9}{\rm Be}^{+} \,}}
\newcommand{\Mg}{\ensuremath{{^{24}}{\rm Mg}^{+} \,}}

\documentclass[prl,onecolumn,floatfix]{revtex4}

\usepackage{graphicx}
\usepackage{amsmath}
\usepackage{amssymb}

\begin{document}

\baselineskip13pt

\title{Complete methods set for scalable ion trap quantum information processing  \footnote{This manuscript has been accepted for publication in
Science.  This version has not undergone final editing.  Please
refer to the complete version of record at
http://www.sciencemag.org/.  The manuscript may not be reproduced or
used in any manner that does not fall within the fair use provisions
of the Copyright Act without the prior, written permission of
AAAS.}}

\author{J.~P.~Home, D.~Hanneke, J.~D.~Jost, J.~M.~Amini, D. ~Leibfried and D.~J.~Wineland.}

\affiliation{\centering Time and Frequency Division, National
Institute of Standards and Technology, Boulder, CO 80305, U.S.A.}

\email{jonathan.home@gmail.com}

\date{\today}

\begin{abstract}
Large-scale quantum information processors must be able to transport
and maintain quantum information, and repeatedly perform logical
operations. Here we demonstrate a combination of all the fundamental
elements required to perform scalable quantum computing using qubits
stored in the internal states of trapped atomic ions. We quantify
the repeatability of a multi-qubit operation, observing no loss of
performance despite qubit transport over macroscopic distances. Key
to these results is the use of different pairs of \Be hyperfine
states for robust qubit storage, readout and gates, and simultaneous
trapping of \Mg ``re-cooling'' ions along with the qubit ions.
\end{abstract}

\maketitle

The long term goal for experimental quantum information processing
is to realize a device involving large numbers of qubits and even
larger numbers of logical operations \cite{05Knill,07Steane}. These
resource requirements are defined both by the algorithms themselves,
and the need for quantum error-correction, which makes use of many
physical systems to store each qubit \cite{03Steane,05Knill}. The
required components for building such a device are robust qubit
storage, single and two-qubit logic gates, state initialization,
readout, and the ability to transfer quantum information between
spatially separated locations in the processor \cite{00DiVincenzo,
07Steane,02Kielpinski}. All of these components must be able to be
performed repeatedly in order to realize a large scale device.

One experimental implementation of quantum information processing
uses qubits stored in the internal states of trapped atomic ions. A
universal set of quantum logic gates has been demonstrated using
laser addressing \cite{08Blatt,03Leibfried,08Benhelm}, leading to a
number of small-scale demonstrations of quantum information
protocols including teleportation, dense-coding, and a single round
of quantum error-correction \cite{08Blatt}. A major challenge for
this implementation is now to integrate scalable techniques required
for large-scale processing.

A possible architecture for a large-scale trapped-ion device
involves moving quantum information around the processor by moving
the ions themselves, where the transport is controlled by time
varying potentials applied to electrodes in a multiple-zone trap
array \cite{98Wineland,02Kielpinski,09Blakestad}. The processor
would consist of a large number of processing regions working in
parallel, with other regions dedicated to qubit storage (memory). A
general prescription for the required operations in a single
processing region is the following (illustrated in Fig. 1), which
includes all of the elements necessary for universal quantum
computation \cite{95Barenco}. (1) Two qubit ions are held in
separate zones, allowing individual addressing for single qubit
gates, state readout, or state initialization. (2) The ions are then
combined in a single zone, and a two-qubit gate is performed. (3)
The ions are separated, and one is moved to another region of the
trap array. (4) A third ion is brought into this processing region
from another part of the device. In this work we implement in a
repeated fashion all of the steps which must be performed in a
single processing region in order to realize this architecture.

\begin{figure}[hb!]
\centering
\includegraphics[scale=1.15]{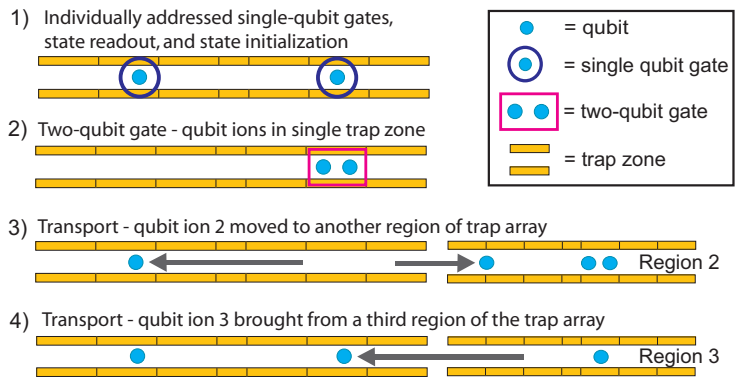}
\caption{ Schematic of the sequence of operations implemented in a
single processing region for building up a computation in the
architecture of \cite{98Wineland,02Kielpinski}. A large-scale device
would involve many of these processing regions performing operations
in parallel, along with additional regions for memory. Generalized
operations would use this block structure repeatedly, with perhaps
some of the steps omitted.}
\end{figure}

Some elements of this architecture have been demonstrated in
previous experiments \cite{04Barrett,08Blatt}, which involved
transport of ions in a multi-zone trap. However, these experiments
did not involve the use of techniques required for building a large
scale device, limiting the size of algorithms which could be
performed. Primary limiting factors for these experiments were the
loss of qubit coherence, caused by interaction with the fluctuating
magnetic field environment, and motional excitation, which degrades
the fidelity of subsequent two-qubit gates because of the finite
wavelength of the gate control fields \cite{00Sorensen1}. Motional
excitation occurs as a result of imperfect control during transport
and noisy electric fields emanating from the electrode surfaces
\cite{08Blatt}. In this work, we store qubits robustly using a pair
of energy eigenstates in the \Be 2s $^2$S$_{1/2}$ hyperfine manifold
(shown in Fig. 2) whose energy separation does not depend on the
magnetic field to first order. For the \Be ``qubit'' ions used here,
this condition is met at a magnetic field of 0.011964 T for the
``memory'' qubit states $\ket{1} \equiv \ket{F = 1, M_F = 0}$ and
$\ket{0} \equiv \ket{F = 2, M_F = 1}$ (The states are labeled using
the total angular momentum quantum numbers $F$ and $M_F$). The
insensitivity to magnetic field changes is crucial for preserving
coherence in the presence of ambient temporal field fluctuations
\cite{05Langer}, and also greatly suppresses phase shifts caused by
spatial variations in the average field experienced by an ion as it
is transported throughout the multi-zone trap array. We remove
motional excitation prior to each two-qubit gate by recooling
``refrigerant'' \Mg ions that are trapped along with the qubit ions.
Laser cooling this second species sympathetically cools the first
through the strong Coulomb interaction between the ions
\cite{03Barrett,01Rohde,02Blinov, 09Home}.

\begin{figure}[ht!]
\centering
\includegraphics[scale=1]{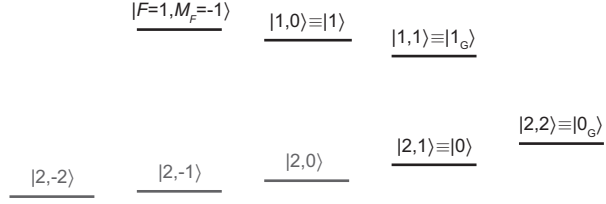}
\caption{ Hybrid qubit storage in the \Be 2s $^2$S$_{1/2}$ hyperfine
levels. The states are labeled using the total angular momentum
quantum numbers $F$ and $M_F$. $\ket{1}$, $\ket{0}$ are the qubit
states used for single qubit gates and transport, and $\ket{1_G}$,
$\ket{0_G}$ are used for two-qubit gates. For detection, the
$\ket{1,-1}$ and $\ket{2,2}$ states are used. At the applied
magnetic field ($B \simeq 0.011964$ T), the frequencies for
transitions between pairs of states with the same $F$ are well
resolved.}
\end{figure}

A benchmark for scalability in this implementation is the repeated
performance of a complete set of one and two-qubit logic gates
combined with quantum information transport. We demonstrate
repeatability of a unitary transformation $\hat{U}$ which involves
four single qubit gates, a two-qubit gate, and transport over
960~$\mu$m (the sequence for $\hat{U}$ is shown in Fig. 3a).
Ideally, $\hat{U}$ implements the operation
\begin{equation}
\hat{U} = -\frac{e^{-i \pi/4}}{\sqrt{2}} \left(\begin{matrix} -1 & 0
& 0 & i \cr 0&1 & i & 0 \cr 0 & i & 1 & 0 \cr i & 0 & 0 & -1
\end{matrix}\right)
\end{equation}
in the $\ket{11}$, $\ket{10}$, $\ket{01}$, $\ket{00}$ basis. We
directly compare experimental implementation of $\hat{U}$ and
$\hat{U}^2$ using quantum process tomography \cite{BkNielsen}.
Process tomography requires the process under investigation to be
applied to sixteen input states, followed by measurement in nine
orthogonal bases \cite{06Riebe}. The input states are prepared using
a combination of optical pumping and single-qubit operations, with
the latter performed on each qubit individually. The analysis also
requires individual single-qubit rotations, followed by individual
state measurement of the qubits. The experiment therefore realizes
all of the basic components illustrated in Fig. 1. We directly
compare $\hat{U}$ and $\hat{U}^2$ by running the experimental
sequence for a given input/output combination on $\hat{U}$ and
$\hat{U}^2$ sequentially (shown in Fig. 3b), making the comparison
of the two robust against long term drifts in experimental
parameters. For each input/analysis combination, we repeat this
sequence 350 times.

\begin{figure}[ht!]
\centering
\includegraphics[scale=1.15]{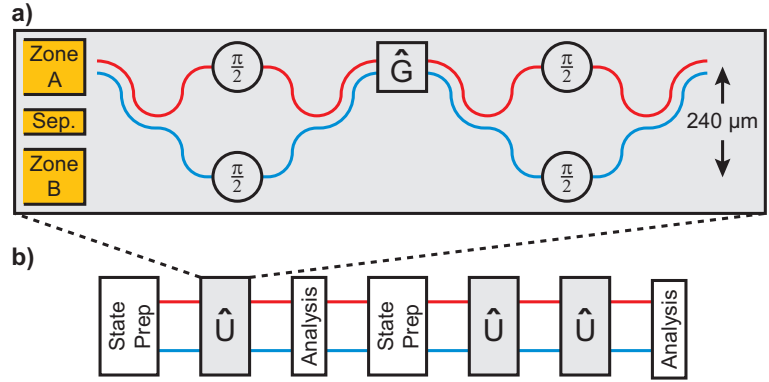}
\caption{ a) Schematic of the qubit ion trajectories (solid red and
dotted blue lines) and gate operations used to implement $\hat{U}$.
The single qubit rotations are ``$\pi/2$'' $\equiv R(\pi/2, 0)$ (eq.
\ref{eq:rotation}). The two-qubit gate implements $\hat{G} =
D[(1,i,i,1)]$. b) Full sequence used to perform process tomography
on $\hat{U}$ and $\hat{U}^2$. This sequence is repeated 350 times
for each setting of preparation/analysis.}
\end{figure}

The experiment utilizes two \Be and two \Mg ions, trapped in a
six-zone linear Paul trap \cite{04Barrett}. Each \Be ion is used to
store one qubit, and is accompanied at all times by a \Mg
refrigerant ion, which is used for sympathetic cooling. The ion
order is initialized to \Be--\Mg--\Mg--\Be at the start of the
experimental sequence, and remains in this order throughout
\cite{09Jost}.

Coherent manipulations of the internal and motional states of the
ions are performed using laser-induced stimulated Raman transitions
\cite{98Wineland}. Single qubit gates are implemented in the basis
$\ket{1}$, $\ket{0}$ by resonant Rabi flopping, applying the
rotation
\begin{equation}
\label{eq:rotation} R(\theta, \phi) = \left(\begin{matrix}
\cos(\theta/2) & -i e^{-i \phi}\sin(\theta/2) \cr -i e^{i \phi}
\sin(\theta/2) & \cos(\theta/2)\end{matrix}\right) \ \ \ ,
\end{equation}
where $\theta$ is proportional to the Raman pulse duration and
$\phi$ is chosen by adjusting the relative phase of the Raman light
fields at the ion. We individually address the two qubit ions by
holding them in two trap zones $240~\mu$m apart, and switching the
laser beams between zones.

To implement two-qubit gates, we first combine all of the ions into
a single zone. The four-ion linear chain exhibits four axial
vibrational normal modes caused by the Coulomb coupling between ions
\cite{09Jost}. After recombination, these modes contain significant
excess energy, mainly caused by imperfect control of the potentials
used during separation and recombination. Therefore, prior to each
two-qubit gate, we cool each mode to near the quantum ground state
($\langle n \rangle \sim 0.06$) using a combination of Doppler
cooling and resolved sideband cooling on the \Mg ions \cite{Methods,
03Barrett}. Importantly, the cooling light only interacts with \Mg,
leaving the qubits stored in \Be intact \cite{03Barrett}.

The composite two-qubit gate makes use of a geometric phase gate
\cite{03Leibfried} to implement $\hat{G} = D[(1, i, i, 1)]$, where
$D[\underline{v}]$ is a diagonal matrix with the vector
$\underline{v}$ on the diagonal. The phase acquired by the
$\ket{10}$ and $\ket{01}$ states is obtained by transient
simultaneous excitation of the two highest-frequency normal modes by
use of a state-dependent optical dipole force \cite{Methods}. The
state dependence of this force is derived from a differential light
shift between the two qubit states, which is highly suppressed for
field-independent transitions \cite{05Lee,05Langer}. We thus use a
hybrid scheme for qubit storage, mapping the qubits into a different
state manifold for the two-qubit gate \cite{Methods,
09Lundblad,09Kirchmair}. Prior to applying the optical dipole force,
we transfer each qubit into a pair of states with a sizeable
differential light shift -- the ``gate'' manifold $\ket{1_G} \equiv
\ket{1, 1}$, $\ket{0_G} \equiv \ket{2, 2}$ (Fig. 2). After applying
the state-dependent force, we reverse this transfer and the ions are
again separated \cite{Methods}. The gate manifold is sensitive to
magnetic field fluctuations, which can lead to qubit dephasing. We
suppress these effects using spin-echo techniques \cite{09Jost}.

We employ quantum process tomography to characterize our
implementation of the unitary operation $\hat{U}$, including any
experimental imperfections \cite{BkNielsen,06Riebe}. The evolution
of the qubit system (including that caused by undesired interactions
with the environment) is described by a completely positive linear
map $\rho_{\rm out} = \mathcal{E}_{\hat{U}}(\rho_{\rm in})$
\cite{BkNielsen} on the input density matrix $\rho_{\rm in} =
\sum_{i,j} c_{i,j} \ket{i}\bra{j}$ , where the $c_{i,j}$ are complex
numbers and $i, j$ are labels that each run over the eigenstates
$\ket{11}$, $\ket{10}$, $\ket{01}$, $\ket{00}$. Following
\cite{BkChpHradil}, we represent the map by a $16\times16$ matrix
\begin{equation}
E_{\hat{U}} = \sum_{i, j} \ket{i}\bra{j} \otimes
\mathcal{E}_{\hat{U}}(\ket{i}\bra{j})\,\,.
\end{equation}
In order to extract this process matrix, we experimentally apply the
process to 16 input states made up of tensor products of the states
$\ket{1}$, $\ket{0}$, $(\ket{0} - \ket{1})/\sqrt{2}$ and $(\ket{0} +
i\ket{1})/\sqrt{2}$. For each output density matrix, we apply nine
sets of rotations, which allow us to measure the expectation values
of the operators $\sigma_s\otimes\sigma_t$, where the $\sigma_{s,
t}$ run over the Pauli matrices $I, \sigma_x, \sigma_y, \sigma_z$.
Our state readout performs a projective measurement in the $Z$ basis
on each ion independently. We first transfer population from
$\ket{0}$ to $\ket{2, 2}$, and from $\ket{1}$ to $\ket{1, -1}$, and
subsequently drive the cycling transition 2s $^2$S$_{1/2} \ket{2,2}
\leftrightarrow $ 2p $^2$P$_{3/2} \ket{3,3}$ for 200~$\mu$s, where
$\ket{2,2}$ strongly fluoresces and $\ket{1, -1}$ does not
\cite{Methods}. We collect a small fraction of the emitted photons
on a photomultiplier tube. We run the sequence shown in Fig. 3b 350
times for each of the 16 input states and nine measurement
rotations. The process matrix is obtained directly from the recorded
photon counts and measurement/preparation settings using a
maximum-likelihood method that ensures that the reconstructed
process matrix is physical \cite{BkChpHradil}.

Experimentally obtained process matrices for one and two
applications of $\hat{U}$ are shown in Fig. 4. From the
reconstructions, we can calculate various measures of the fidelity
with which the processes were implemented. A direct comparison
between experimental results and the ideal case is given by the
entanglement fidelity $F \equiv {\rm Tr}(E_{\rm ideal} E)/16$
\cite{99Horodecki}. We find $F_{\hat{U}} = 0.922(4)$ for a single
application of $\hat{U}$, and $F_{\hat{U}^2} = 0.853(5)$ for two
applications (error estimates are the standard error on the mean
obtained from parametric bootstrap resampling \cite{Methods}). As an
additional measure of operation fidelity, we take the mean $\bar{f}$
of the fidelity $f(\rho_{\rm ideal}, \rho_E) \equiv \left[{\rm
Tr}(\sqrt{\sqrt{\rho_{\rm ideal}} \rho_E \sqrt{\rho_{\rm
ideal}}})\right]^2$ \cite{94Jozsa} between the output density
matrices obtained from the ideal and experimental processes for an
unbiased set of 36 input states (formed from the eigenstates of
$\sigma_s\otimes\sigma_t$, where $\sigma_{s,t}$ run over $\sigma_x,
\sigma_y, \sigma_z$). We obtain a mean state fidelity of
$\bar{f}_{\hat{U}} = 0.940(4)$ for $E_{\hat{U}}$ and
$\bar{f}_{\hat{U}^2} = 0.890(4)$ for $E_{\hat{U}^2}$. We can compare
these values to the entanglement fidelities using the relation
$\bar{f} = (4 F + 1)/5$ \cite{99Horodecki}, and see that they are
consistent.

\begin{figure}[ht!]
\centering
\includegraphics[scale=.5]{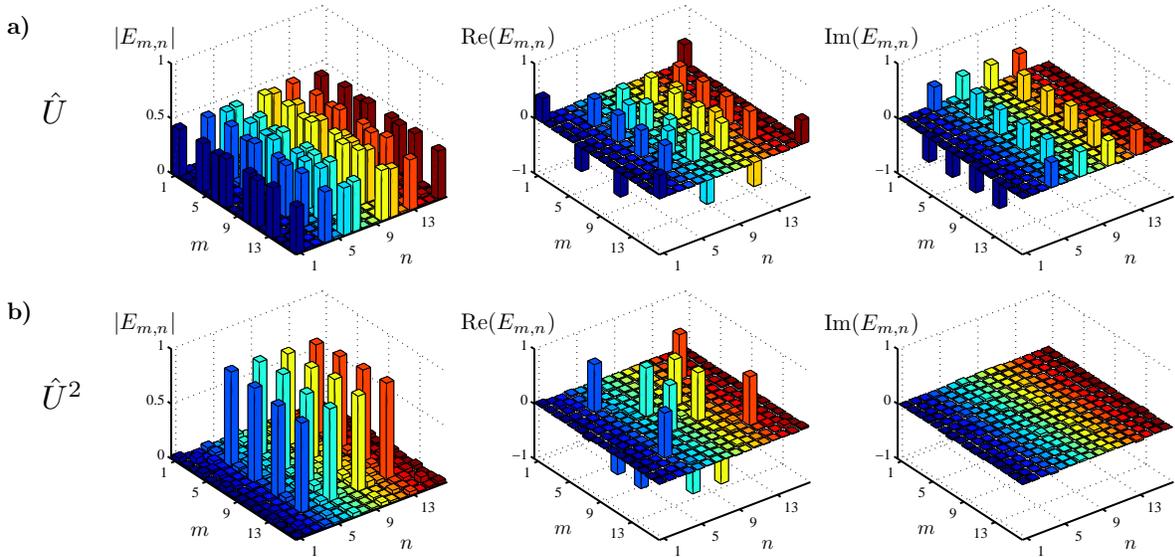}
\caption{  Reconstructed process matrix for a) $\hat{U}$ and b) two
repetitions of $\hat{U}$. The map $\mathcal{E}(\ket{i}\bra{j})$
produces a matrix $\mathcal{E}_{k,l}$ for each element
$\ket{i}\bra{j}$. Hence elements of the matrix $E$ are labeled by $m
= 4 (i - 1) + k, n = 4 (j - 1) + l$, where the factor 4 results from
the size of the two-qubit state space. For example, the
$\ket{11}\bra{00}$ ($i = 1$, $j = 4$) element of an input density
matrix is mapped to $\mathcal{E}(\ket{11}\bra{00})$, a 4$\times$4
block of $E$ given by $m \in [1,4]$ and $n \in [13,16]$. The
position of each peak is in agreement with the theoretical
prediction.}
\end{figure}

To compare the performance of a second application of $\hat{U}$
relative to the first, we can compare its experimental repetition
$\mathcal{E}_{\hat{U}^2}(\rho_{\rm in})$ to a numeric repetition of
the experimental map $\mathcal{E}_{\hat{U}}(\rho_{\rm in})$, i.e.,
to $\mathcal{E}_{\hat{U}\hat{U}}(\rho_{\rm in}) \equiv
\mathcal{E}_{\hat{U}}(\mathcal{E}_{\hat{U}}(\rho_{\rm in}))$.
Evaluating the fidelities for each against the ideal case yields
$F_{\hat{U}^2}/F_{\hat{U}\hat{U}} = 1.003(13)$ and
$\bar{f}_{\hat{U}^2}/\bar{f}_{\hat{U}\hat{U}} = 1.004(10)$,
indicating that the operation fidelity is the same for each
application of $\hat{U}$. We can also make a direct comparison
between the processes performed by our implementation of $\hat{U}$
and $\hat{U}^2$ by taking the mean fidelity between $\rho_{\rm
\hat{U}\hat{U}} = \mathcal{E}_{\hat{U}}
(\mathcal{E}_{\hat{U}}(\rho_{\rm in}))$ and $\rho_{\rm \hat{U}^2} =
\mathcal{E}_{\hat{U}^2}(\rho_{\rm in})$ for the 36 input states. We
find $\bar{f}(\rho_{\rm \hat{U}\hat{U}}, \rho_{\rm \hat{U}^2}) =
0.987(3)$. Although this number is not unity, as might be expected,
the deviation can be ascribed to bias in the maximum-likelihood
reconstruction method for finite sample size \cite{Methods}. Our
results are thus consistent with the same operation being performed
by the experiment for each application of $\hat{U}$.

Sources of error in our system arise primarily from spontaneous
photon scattering ($\sim$1.5\% per $\hat{U}$) \cite{07Ozeri} and
intensity fluctuations of the Raman light fields at the percent
level. In order to characterize the loss of fidelity caused by
single-qubit rotations, we apply process tomography to the
experimental sequence, but without the state-dependent force pulses.
In this case the ions are always in a product state and the process
matrix for each can be obtained independently. The resulting process
matrices have mean state fidelity relative to the ideal case of 0.97
for a single run of the sequence (which uses eight rotations per ion
including qubit manifold transfer and spin-echo pulses). During the
two-qubit gate, the spin states are entangled with the motion. From
separate measurements of motional coherence, we estimate the
infidelity from this source to be less than $1\times 10^{-3}$.

Many challenges remain before large-scale ion trap quantum
information processing becomes a reality, including increasing
fidelities to those required for fault-tolerant quantum error
correction \cite{03Steane,05Knill}, and meeting the considerable
technical challenge of controlling ions in large multi-dimensional
trap arrays \cite{09Blakestad}. Both of these challenges could
potentially contain problems which have not been considered here,
and which may require combining our approach with alternative
methods, for instance entanglement distribution using photonic
networks \cite{07Moehring}. Nevertheless, the combination of
techniques demonstrated here includes all of the basic building
blocks required in this architecture, and opens up new possibilities
for quantum information processing as well as state and process
engineering.

 This work was supported by IARPA and the NIST Quantum Information
Program.
 J.~P.~H. acknowledges support from a Lindemann Trust Fellowship.
 We thank E. Knill for helpful discussions, J. J. Bollinger for technical
 assistance and Y.~Colombe for comments on the manuscript. This paper is a
 contribution by the National
 Institute of Standards and Technology and not subject to U.S. copyright.


\section*{Materials and Methods}

The Supporting Online Materials provide further details about the
sympathetic cooling, the use of state-dependent forces to implement
a geometric phase gate using multiple motional modes, transfer
between the gate and memory qubit manifolds, state detection and
error analysis for quantum process tomography.

\subsection{Sympathetic cooling.}
After recombining the ions in a single trap zone, each of the axial
modes of the four ion \Be-\Mg-\Mg-\Be chain contains significant
excess energy. The precise amount is difficult to characterize,
since the distribution of population in the Fock state basis is not
well known. For the experiments presented here, we require three
stages of cooling to attain the ground state. We start with Doppler
cooling, which thermalizes the state with a mean vibrational
occupation number $\bar{n} \sim 15$. We then use cycles of resolved
sideband cooling on both the second and first sidebands of each
motional mode to prepare the ground state with a fidelity of $\sim
0.94$ for each mode. In total, the cooling takes 5.1~ms per
$\hat{U}$, and this is the limiting factor in the operation time
(followed by transport/separation, which takes 3.6~ms per
$\hat{U}$). The motional excitation produced in our experiment is
much higher than in similar previous work \cite{04Barrett}, which we
attribute primarily to use of voltage supplies with an voltage
update rate below the typical trap frequencies \cite{09Blakestad}.
In future this time should be able to be substantially reduced by
using supplies with an increased update rate.

\subsection{Two-qubit logic gate.}
The two-qubit gate follows the basic method used in
\cite{03Leibfried}, with the modification that we simultaneously
excite two motional modes rather than one. This speeds up the gate
for a given laser intensity, and thus reduces the error due to
spontaneous photon scattering. We perform the gate with the ions in
the spatial configuration \Be-\Mg-\Mg-\Be. The two modes we excite
are the highest frequency axial modes, which are separated in
frequency by $2 \pi \times 251$~kHz \cite{09Jost}. The difference
frequency of the two Raman light fields that generate the
state-dependent force is $\omega = \omega_3 + \delta = \omega_4 - 2
\delta$, where $\omega_3, \omega_4$ are the frequencies of the
motional modes, and $\delta = 2 \pi \times 83.6$~kHz. The
state-dependent force is applied for a duration $t_G = 2
\pi/\delta$, ensuring that the spin states are disentangled from
both motional modes at the end of the pulse \cite{03Leibfried}. We
choose the axial confinement such that the two $\Be$ ions are
separated by a half-integer number of wavelengths of the optical
dipole potential. The oscillation of the two \Be ions in the
$\omega_4$ mode is out of phase, resulting in motional excitation of
this mode for $\ket{0_G 1_G}$ and $\ket{1_G 0_G}$. For the
$\omega_3$ mode, motion is excited (and thus phase acquired) for all
qubit states. This can lead to unwanted $\hat{\sigma}_z$ rotations
being implemented along with the phase gate. For this reason, and to
increase robustness against other sources of error (such as Stark
shifts) during the motional excitation, we apply the phase gate in
two pulses, with a $R(\pi, \phi_G)$ pulse on $\ket{0_G}
\leftrightarrow \ket{1_G}$ applied between the two. This is followed
by a second $R(\theta, \phi_G')$ pulse (see next section). The ratio
of contributions to the gate phase due to each mode is
$\Phi_3/\Phi_4 = 2.72$.

\subsection{Transfer between the gate and memory
qubit manifolds}

In order to implement the state-dependent optical dipole force, we
transfer from the field-independent manifold  $\ket{1}, \ket{0}$ to
the ``gate'' manifold $\ket{1_G}, \ket{0_G}$. This is performed by
applying $R(\pi, \phi)$ to $\ket{1}\leftrightarrow \ket{1_G}$ and
$\ket{0} \leftrightarrow \ket{0_G}$ sequentially (for these and
other transitions described in this section, the rotation basis is
given by ($\ket{h}$, $\ket{l}$), where $h$ ($l$) indicates the
higher (lower) energy state). Each of these pulses imprints a
different phase $\phi_1, \phi_0$ on the qubit state. After the
state-dependent force pulses have been applied, we repeat this
transfer in reverse order, taking the qubit back into the $\ket{1},
\ket{0}$ manifold and imprinting phase $\phi_1', \phi_0'$. While in
the gate manifold, we apply two spin-echo pulses $R(\pi,
\phi_G),R(\pi, \phi_G')$ on $\ket{1_G} \leftrightarrow \ket{0_G}$.
As a result the final phase imprinted on the qubit by these
rotations depends on the phase differences $\phi_1 - \phi_1'$,
$\phi_0 - \phi_0'$ and $\phi_G - \phi_G'$. If each transition is
driven with independent oscillators these phase differences are
zero. In our case these are non-zero due to the use of tunable
frequency sources for driving Raman transitions, and we adjust the
phase of all subsequent single-qubit gates to compensate for this.

\subsection{Detection.}
The state of each ion is read out individually. Due to the lack of
multiple detection apparatus, we perform state readout sequentially.
First, with the ions in separate zones, we transfer population from
the qubit memory levels into the qubit readout levels $\ket{2,2}$
and $\ket{1, -1}$, using a sequence of $R(\pi, 0)$ pulses on
$\ket{2, 1} \rightarrow \ket{2,2}$ and
$\ket{1,0}\rightarrow\ket{1,-1}$. We then resonantly drive the
cycling transition 2s $^2$S$_{1/2} \ket{2,2} \leftrightarrow $ 3p
$^2$P$_{3/2} \ket{3,3}$ for the ion in trap zone A (ion 1) and
detect. For an ion in $\ket{2,2}$, the distribution of the number of
photons collected in the 200~$\mu$s detection period closely
approximates a Poisson distribution with mean 10 counts. For an ion
in $\ket{1, -1}$, the distribution has two components, the dominant
contribution coming from background scatter of the laser light
(which has a Poisson distribution with mean ~0.4 counts), and a much
smaller exponential contribution due to repumping of the $\ket{1,
-1}$ state into $\ket{2, 2}$ over the detection period (the
repumping probability is $\approx 1\times10^{-3}$ over 200~$\mu$s).
In order to detect the second ion, we recombine the ions in zone A,
and again drive the cycling transition. To ensure that ion 1
contributes negligibly to the second detection, prior to
recombination we transfer all population in this ion to $\ket{1,
-1}$ by optical pumping to $\ket{2,2}$ followed by $R(\pi, 0)$
transfer pulses on $\ket{2,2}\rightarrow\ket{1,1}$,
$\ket{1,1}\rightarrow\ket{1,0}$, $\ket{1,0}\rightarrow\ket{1,-1}$.
If the first of these was imperfect, it might leave population in
the fluorescing state, hence we perform a transfer pulse from
$\ket{2,2}\rightarrow\ket{2,1}$ to protect against this ($\ket{2,1}$
scatters relatively few photons for our detection parameters).

\subsection{Tomography error analysis by resampling.}
Statistical errors for values given in the text are derived from
parametric bootstrap resampling \cite{BkEfron}. Using the process
matrix obtained from the experimental data, we generated 100 sample
data sets. This was performed using a random number generator
combined with the known probability distributions for the photon
counts, and choosing from a normal distribution of Rabi frequencies
consistent with observed intensity fluctuations for the rotation
angles $\theta$ used for preparation/analysis rotations. For each
sample data set, we obtained a process matrix, and evaluated the
relevant fidelities. The quoted statistical error estimates are the
standard error on the mean obtained from the distributions of
results. These distributions are centred about one standard error
below the value obtained directly from the experimental data. This
shift is due to the fact that the maximum-likelihood estimation is
only an unbiased estimator in the limit of infinite sample size, and
that the estimation constrains the solution to lie within the set of
physically allowed process matrices. This implies that the quoted
experimental fidelities are underestimates.

The effect of reconstruction bias is illustrated by the following
example. We numerically generated 100 sample data sets using both
the experimentally obtained $\mathcal{E}_{\hat{U}}(\rho)$ and the
application of this same map twice
$\mathcal{E}_{\hat{U}}(\mathcal{E}_{\hat{U}}(\rho))$. For each data
set, we performed the maximum likelihood analysis, obtaining
$E_{\hat{U}}'$ and $E_{\hat{U}^2}'$. We compared the performance of
these two processes by taking the mean fidelity between
$\rho_{\hat{U}\hat{U}}' =
\mathcal{E}_{\hat{U}}'(\mathcal{E}_{\hat{U}}'(\rho))$ and
$\rho_{\hat{U}^2}' = \mathcal{E}_{\hat{U}^2}'(\rho)$ over the set of
36 input states. The average value of this fidelity over the 100
sample data sets was 0.989(2), where for unbiased reconstruction
this value would be one. This means that the experimental value of
0.987(3) is consistent with the gate operation being the same for
each repetition.

\end{document}